\documentclass[12pt]{article}
\usepackage{amssymb}
\usepackage{latexsym}
\newcommand{\Z}{{\mathbb Z}}

\def\be{\begin{equation}\label}
\def\ee{\end{equation}}
\def\bea{\begin{eqnarray}\label}
\def\eea{\end{eqnarray}}
\def\Tr{{\rm \,Tr\,}}
\def\d{{\rm d}}

\begin{document}
\title{{\flushleft{\small {\rm Journal reference: J. Phys. A: Math. Gen. \textbf{35} (2002) 6995-7002}\\
Addenda after Eqs.~(34) and(44)\\
\vspace{2cm}}}
\large\bf Percolation transition in the Bose gas II.}
\author{Andr\'as S\"ut\H o\\
Research Institute for Solid State Physics and Optics\\ Hungarian Academy of
Sciences \\ P. O. Box 49, H-1525 Budapest\\ Hungary\\
E-mail: suto@szfki.hu}
\date{}
\maketitle
\thispagestyle{empty}
\begin{abstract}
\noindent
In an earlier paper (J. Phys. A: Math. Gen. {\bf 26} (1993) 4689)
we introduced the notion of cycle percolation in the Bose gas and
conjectured that it occurs if and only if there is Bose-Einstein condensation. Here we
give a complete proof of this statement for the perfect and the imperfect (mean-field)
Bose gas and also show that in the condensate there is an infinite number of
macroscopic cycles.
\vspace{5mm}\\
PACS: 0530J, 0550, 7510J
\vspace{1cm}\\
\end{abstract}
\newpage

\section{Introduction}

It was perhaps Feynman who first emphasised the importance of long permutation cycles
in the description of the $\lambda$-transition in liquid helium and, more generally, of
Bose-Einstein condensation (Feynman 1953).
Some years ago the present author gave a mathematically precise formulation of this idea
by introducing the notion of cycle percolation in Bose systems
(S\"ut\H o 1993, hereafter referred to as I).
In I it was suggested that this concept could
serve as an alternative characterization of Bose-Einstein condensation (BEC). The examples of the ground
state of a system of bosons and the three dimensional ideal Bose gas were treated. For the
latter it was shown that BEC indeed implies cycle percolation, but the proof of the absence of
percolation in the absence of BEC was missing. Since then, this kind of description
has gained some new interest (e.g. Bund and Schakel 1999, Schakel 2000, Martin 2001, Ueltschi 2002).
Therefore, it seems to be worthwhile to complete the earlier proof and to make the claim stronger by
showing the occurrence of an infinity of macroscopic cycles in the condensate. As it will become
clear, our conclusions also apply to the imperfect Bose gas.

Let us recall the principal definitions and relevant results of I. Let $\Lambda$ be a
cube of side $L$ of the $d$ dimensional Euclidean space, and denote $H_{\Lambda,N}$ the Hamiltonian
of $N$ (interacting) bosons in $\Lambda$, taken, e.g., with periodic boundary conditions.
By making explicit the summation over
the permutations of $N$ particles, necessary for symmetrization,
the corresponding canonical partition function reads
\bea{part}
Q_{\Lambda,N}=\Tr_{{\mathcal H}_{\rm sym}^N}e^{-\beta H_{\Lambda,N}}
={1\over N!}\sum_{g\in S_N}\Tr_{{\mathcal H}^N}U(g)e^{-\beta H_{\Lambda,N}}\nonumber\\
=\sum_{\{n_j\}:\sum jn_j=N}\left(\prod_{j=1}^N\frac{1}{j^{n_j}n_j!}\right)
\Tr_{{\mathcal H}^N}U(g_{\{n_j\}})e^{-\beta H_{\Lambda,N}}
\eea
Here $S_N$ is the group of permutations of $\{1,2,\ldots,N\}$ and $U(g)$ is the unitary representation of $g$
in the $N$-particle Hilbert space ${\mathcal H}^N$ where ${\mathcal H}=L_2(\Lambda)$.
In the second line of (\ref{part}) the summation goes over the conjugation classes of $S_N$. Each $g$
in the same class giving the same contribution, see (I.2.5), $g_{\{n_j\}}$ can be any permutation with
$n_j$ cycles of length $j$.
Because each term contributing to the sum is positive, cf equation (I.2.18),
\be{prob}
P_{\Lambda,N}(g)={1\over N!\,Q_{\Lambda,N}}\Tr_{{\mathcal H}^N}U(g)e^{-\beta H_{\Lambda,N}}
\ee
can be interpreted as the probability of the occurrence of the permutation $g$ in the canonical
ensemble. Then, we can introduce random variables, for example $\xi_1$, the length of the cycle
containing 1 (i.e. particle no 1) and $n_j$, the number of cycles of length $j$.
Constancy of the probability (\ref{prob}) within a conjugation class means
that the probability is sensitive only to the number and length of the cycles and not to their contents.
Thus,
1 is assigned to any cycle with a probability proportional to the length of the cycle, which yields
the conditional probability
\be{pcond}
P_{\Lambda,N}(\xi_1=j|n_j=m)=mj/N
\ee
and, as a consequence,
\be{j}
P_{\Lambda,N}(\xi_1=j)=\frac{j}{N}\sum_m m P_{\Lambda,N}(n_j=m)\equiv {j\langle n_j\rangle_{\Lambda,N}\over N}\ .
\ee
Clearly, $\sum_{j=1}^N P_{\Lambda,N}(\xi_1=j)=1$. However, if we first perform the thermodynamic
limit $L\to\infty$, $N\to\infty$ in such a way that $N/L^d\to\rho>0$,
and then sum over the limiting probabilities $P_\rho(\xi_1=j)$, we can state only that
\be{sumPrho}
\sum_{j=1}^\infty P_\rho(\xi_1=j)\leq 1\ .
\ee
We speak about cycle percolation, in an obvious analogy with site or bond percolation, if in this
relation the strict inequality holds. In I we conjectured that this happens if and only if there
is BEC. Half of the conjecture, namely that BEC implies cycle percolation was shown for the
free Bose gas. Now we give a complete proof of a somewhat stronger assertion.

In the theorem
below by {\it condensate} we mean the ensemble of particles in the one-particle ground state, when
BEC occurs. Throughout the paper we work with periodic boundary conditions, so these are particles
with zero wavevector. Furthermore, a {\it macroscopic cycle} is a cycle of positive density, i.e.
a cycle containing a non-vanishing fraction of the total number of particles.

\vspace{5pt}
\noindent
{\bf Theorem} {\it In the perfect and mean-field Bose gases Bose-Einstein condensation occurs
if and only if there is cycle percolation. In the condensate the number of macroscopic cycles
is (countable) infinite. Moreover, there are no finite cycles in the condensate and no macroscopic
cycles outside it.}

\vspace{5pt}
In the following section we prove the theorem in three steps. First, we recall the proof presented in I
for the existence of cycle percolation when there is condensation. The second and third parts contain
the new results: the absence of percolation in the absence of condensation and the analysis of the
cycles in the condensate.

\section{Proof of the Theorem}
\subsection{Proof of cycle percolation in case of condensation}

For the ideal Bose gas
\be{freeHam}
H_{\Lambda,N}=T_{\Lambda,N}=-{\hbar^2\over 2m}\sum_{i=1}^N\Delta_i\ ,
\ee
the N-particle kinetic energy operator with, as we suppose in this paper,
periodic boundary conditions in $\Lambda$. In rewriting (\ref{part}), we make use
of the fact that $U$ is a representation of $S_N$ and, therefore,  $U(g_{\{n_j\}})$ can be
decomposed into a product of $\sum n_j$ commuting factors, each corresponding to a cycle.
Because now $\exp(-\beta T_{\Lambda,N})$ also factorizes, the trace will factorize according to the
cycles, and
the partition function reads
\be{freepart}
Q_{\Lambda,N}=\sum_{\{n_j\}:\sum jn_j=N}\prod_{j=1}^N\frac{1}{n_j!}\left(\frac{1}{j}
\Tr e^{-j\beta T_{\Lambda,1}}\right)^{n_j}
\ee
with the trace over ${\mathcal H}$.
The key to the proof is provided by the bounds
\be{bounds}
\left(\frac{L}{\lambda_B\sqrt{j}}-1\right)^d<\Tr e^{-j\beta T_{\Lambda,1}}
<\left(\frac{L}{\lambda_B\sqrt{j}}+1\right)^d\ .
\ee
These are direct consequences of the inequalities
\be{ineq}
\frac{1}{2}\sqrt{\pi\over\alpha}-1<\int_1^\infty e^{-\alpha x^2}\d x
<\sum_{n=1}^\infty e^{-\alpha n^2}<\int_0^\infty e^{-\alpha x^2}\d x
=\frac{1}{2}\sqrt{\pi\over\alpha}
\ee
applied with $\alpha=\pi j\lambda_B^2/L^2$ where $\lambda_B=\hbar\sqrt{2\pi\beta/m}$, the thermal
de Broglie wavelength. Now equation (\ref{j}) with (\ref{freepart}) yields
\be{freej}
P_{\Lambda,N}(\xi_1=j)=\frac{1}{N}{Q_{\Lambda,N-j}\over Q_{\Lambda,N}}\Tr e^{-j\beta T_{\Lambda,1}}
<\frac{1}{N}\Tr e^{-j\beta T_{\Lambda,1}}\ .
\ee
The inequality holds because $Q_{\Lambda,N-j}<Q_{\Lambda,N}$ for periodic boundary conditions.
(See (I.6.16) for Dirichlet boundary condition.)
Taking the thermodynamic limit and using the upper bound in (\ref{bounds}),
we find
\be{freePrho}
P_\rho(\xi_1=j)\leq \frac{1}{j^{d/2}\rho\lambda_B^d} \ .
\ee
This leads to
\be{freesumPrho}
\sum_{j=1}^\infty P_\rho(\xi_1=j)\leq{1\over\rho\lambda_B^d}\sum_{j=1}^\infty {1\over j^{d/2}}=
{g_{d/2}(1)\over\rho\lambda_B^d}
\ee
where
\be{g}
g_\alpha(z)=\sum_{j=1}^\infty{z^j\over j^\alpha}\ .
\ee
According to the well-known condition (e.g. Huang 1987), there is BEC in the free Bose gas if and only if
the right-hand side of (\ref{freesumPrho}) is less than 1, which implies cycle percolation.

\subsection{Proof of the absense of percolation if there is no condensation}

Here we refer to the strong
equivalence of ensembles in the absence of BEC. As we show below, for
any fugacity $z<1$ and any fixed
$j$  the grand-canonical probability distribution of $n_j/L^d$ becomes degenerate in the
thermodynamic limit, i.e. will be concentrated onto a single value. As a consequence, the canonical
probability distribution of $n_j/L^d$ will also be degenerate and concentrated on the same value, provided
that the particle density $\rho$ corresponds to $z$.
So it will be possible to compute $P_\rho$ from the grand-canonical probabilities $P_z$.
First we show the absence of
cycle percolation in the absence of BEC in the grand-canonical ensemble.

The grand-canonical partition function
\bea{grand}
Z_{\Lambda,z}&=&\sum_{N=0}^\infty z^N Q_{\Lambda,N}=\sum_{N=0}^\infty \frac{z^N}{N!}\sum_{g\in S_N}
\Tr_{{\mathcal H}^N}U_N(g)e^{-\beta H_{\Lambda,N}}\\
&=&\sum_{\{n_j\}_{j\geq 1}:\sum n_j<\infty}\left(\prod_{j}\left(\frac{z^j}{j}\right)^{n_j}
\frac{1}{n_j!}\right)
\Tr_{{\mathcal H}^{\sum jn_j}}U_{\sum jn_j}(g_{\{n_j\}})e^{-\beta H_{\Lambda,\sum jn_j}}\nonumber
\eea
can also be considered as the generator of a probability distribution over $\cup S_N$. Since any $g\in S_N$
is naturally embedded into $S_{N'}$ for any $N'>N$ (by setting $g(i)=i$ for $i>N$), it is more precise
to say that now probabilities are
assigned to the {\it representations} $U_N(g)$,
\be{gcprob}
P_{\Lambda,z}(U_N(g))=\frac{z^N}{N!Z_{\Lambda,z}}\Tr_{{\mathcal H}^N}U_N(g)e^{-\beta H_{\Lambda,N}}
\ee
or, as the second line of (\ref{grand}) shows it, to terminating sequences
of non-negative integers, each $\{n_j\}$ with $\sum n_j<\infty$
corresponding to a conjugation classe of $S_N$ for $N=\sum jn_j$.
As in the case of the canonical ensemble, we will consider $n_j$ to be random
variables. For a general Hamiltonian any finite number of them has a coupled joint distribution.
However, if for the ideal Bose gas we replace (\ref{freepart}) into equation (\ref{grand}),
we arrive at
\be{freegrand}
Z_{\Lambda,z}=
\prod_{j=1}^\infty \exp\left\{\frac{z^j}{j}\Tr e^{-j\beta T_{\Lambda,1}}\right\}\ ,
\ee
showing that the $n_j$ are independent random variables with probabilities
\be{gcP}
P_{\Lambda,z}(n_j=n)=\exp\left\{-\frac{z^j}{j}\Tr e^{-j\beta T_{\Lambda,1}}\right\}\frac{1}{n!}
\left(\frac{z^j}{j}\Tr e^{-j\beta T_{\Lambda,1}}\right)^n\ .
\ee
The $L$-dependence of (\ref{gcP}) leads to the asymptotic degeneracy of the grand-canonical
distribution of $n_j/L^d$ for $z<1$:
\be{asy}
\lim_{L\to\infty}P_{\Lambda,z}(\frac{n_j}{L^d}\leq x)=\Theta(x-\rho_j)
\ee
where $\Theta$ is the Heaviside function and
\be{rhoj}
\rho_j=\frac{z^j}{j^{1+d/2}\lambda_B^d}\ .
\ee
Equation (\ref{asy}) is the manifestation of the following large deviation principle.
Let $0<a<b$, then
\be{LDP}
\lim_{L\to\infty}\frac{1}{L^d}\ln P_{\Lambda,z}(a\leq {n_j\over L^d}\leq b)=-\inf_{a\leq x\leq b}I_j(x)\ .
\ee
where the rate function $I_j$ is given by
\be{rate}
I_j(x)=\rho_j-x+x\ln\frac{x}{\rho_j}\ .
\ee
To prove (\ref{LDP}) and (\ref{asy}), we extend the right-hand-side of (\ref{gcP}) to real
positive values of $n$ and define
\be{fl}
f_L(x)=\exp\left\{-\frac{z^j}{j}\Tr e^{-j\beta T_{\Lambda,1}}\right\}\frac{1}{(xL^d)!}
\left(\frac{z^j}{j}\Tr e^{-j\beta T_{\Lambda,1}}\right)^{xL^d}
\ee
for $x>0$.
Then, by using (\ref{bounds}) and Stirling's formula, asymptotically,
as $L\to\infty$, we find
\be{flasy}
f_L(x)\asymp \frac{1}{\sqrt{2\pi xL^d}}e^{-L^d I_j(x)}\ .
\ee
Equations (\ref{asy}) and (\ref{LDP}) follow from (\ref{flasy}) by noting that $I_j(x)$ is strictly
convex for $x>0$ and $I_j(\rho_j)=0$ is its minimum. For the distribution of $n_j$ this implies that
asymptotically $n_j$ is concentrated to an $O(L^{d/2})$-neighbourhood of $\rho_jL^d$.

As in the canonical case, $\xi_1$ is the length of the cycle containing 1. In finite volumes
\bea{pxi1}
P_{\Lambda,z}(\xi_1=j)&=&\sum_{\{m_i\}:\sum m_i<\infty}P_{\Lambda,z}(\xi_1=j|n_i=m_i,i\geq 1)
P_{\Lambda,z}(n_i=m_i,i\geq 1)\nonumber\\
&=&\sum_{\{m_i\}:\sum m_i<\infty}\frac{jm_j/L^d}{\sum im_i/L^d}
\prod_i P_{\Lambda,z}(n_i/L^d=m_i/L^d)\ .
\eea
With our large-deviation result then
\be{Pzxi}
P_z(\xi_1=j)=\lim_{L\to\infty}P_{\Lambda,z}(\xi_1=j)=\frac{j\rho_j}{\sum_i i\rho_i}=
{z^j\over j^{d/2}g_{d/2}(z)}\ .
\ee
Since the denominator is finite for $z<1$, we find $\sum_{j=1}^\infty P_z(\xi_1=j)=1$
indeed. Now for $z<1$ $\rho$ and $z$ are related through
\be{dens}
g_{d/2}(z)=\rho\lambda_B^d\ .
\ee
Substituting this expression into (\ref{Pzxi}) we obtain $P_\rho$ and the absence of cycle percolation in the
canonical ensemble.

Two remarks are in order here:

1. As we have mentioned at the beginning of this section, the asymptotic degeneracy (\ref{asy}) implies
that the canonical distribution of $n_j/L^d$ is also asymptotically
degenerate. Knowing this, we can derive $P_\rho(\xi_1=j)$ without passing through $P_z(\xi_1=j)$.
Indeed, in equation (\ref{j})
for $j=1,\ldots,N$ the averages $\langle n_j\rangle_{\Lambda,N}$ can be approximated by the set $\{n_j\}$
belonging to the largest term of (\ref{freepart}), which can be found by conditional maximization.
The asymptotically valid result is
\be{njmax}
n_j\asymp {N z_N^j\over j^{1+d/2}\rho\lambda_B^d}
\ee
Here $z_N$ appears via a Lagrange multiplier and satisfies
\be{zN}
\sum_{j=1}^N{z_N^j\over j^{d/2}}=\rho\lambda_B^d\ .
\ee
If $\rho<\rho_c(\beta)=g_{d/2}(1)/\lambda_B^d$ then $z_N<1$. Writing equation (\ref{zN})
for $N$ and $N+1$ and taking the difference one can see that for any fixed
$\rho>0$, $z_N>z_{N+1}$. So for $\rho<\rho_c$, $z_N\downarrow z<1$
which, hence, is just the solution for $z$ of equation (\ref{dens}). In sum,
\be{Papprox}
P_{\Lambda,N}(\xi_1=j)\asymp {z_N^j\over j^{d/2}\rho\lambda_B^d}
\to P_\rho(\xi_1=j)=\frac{z^j}{j^{d/2}\rho\lambda_B^d}\ .
\ee

2. Comparing the expression (\ref{Pzxi}) of $P_{z=1}(\xi_1=j)$ with the right member of the inequality
(\ref{freePrho}), we see that the inequality saturates at $\rho=\rho_c$. In fact, for $\rho\geq\rho_c$
there is equality in (\ref{freePrho}): Although for $\rho>\rho_c$ there is no asymptotic equivalence
between the canonical and grand canonical distributions of $n_j/L^d$
(reflecting the fact that the canonical and grand canonical Gibbs states are related through the
nondegenerate Kac density, see Cannon 1973 and Lewis and Pul\`e 1974),
the former still becomes asymptotically
degenerate. So the strong equivalence of the canonical ensemble with a microcanonical one (with
$n_j$ fixed) persists. The largest term of (\ref{freepart}) is given by (\ref{njmax}). From
(\ref{zN}) it follows that $z_N>1$ and converges to 1, therefore
\be{Prho=}
P_\rho(\xi_1=j)=\frac{1}{j^{d/2}\rho\lambda_B^d}\quad{\rm if}\quad\rho>\rho_c\ .
\ee

\subsection{Existence of an infinity of macroscopic cycles}

\vspace{2pt}
While $\sum_{j=1}^\infty P_\rho(\xi_1=j)<1$ means that infinite cycles occur with positive
probability, this inequality does not inform us about the asymptotic size and number of large
cycles. Below we show that in the domain of condensation macroscopic cycles are in
abundance, their number is infinite, and each of them is associated with the condensate.

Let, therefore, $d\geq 3$ and fix a $\rho>\rho_c(\beta)=g_{d/2}(1)/\lambda_B^d$.
Thus, $\rho_0\equiv\rho-\rho_c>0$. Let, moreover, $N=N(L,\rho)=[\rho L^d]$. Writing the trace
in the basis of the eigenstates of $T_{\Lambda,N}$, the partition function reads
\be{frpart}
Q_{\Lambda,N}={1\over N!}\sum_K e^{-\beta\sum_k N_k(K)\varepsilon_k}\sum_g\delta_{gK,K}\ .
\ee
Here $K=(k_1,\ldots,k_N)$, $k_i\in\Lambda^*=(2\pi/L)\Z^d$, $\varepsilon_k=(\hbar^2k^2/2m)$,
$N_k(K)$ is the number of occurrence
of $k$ in the sequence $K$ and $gK=(k_{g^{-1}(i)})_{i=1}^N$. From (\ref{frpart}) we can infer the
probability of a pair $(g,K)$,
\be{probkg}
P_{\Lambda,N}(g,K)={e^{-\beta\sum_k N_k(K)\varepsilon_k}\delta_{gK,K}\over N!\, Q_{\Lambda,N}}
\ee
and that of $K$, $P_{\Lambda,N}(K)=\sum_g P_{\Lambda,N}(g,K)$.

The asymptotic distribution of the random variables $N_k$ was studied in detail by Buffet and Pul\`e (1983).
They showed that the canonical distribution of $N_k/L^d$ becomes asymptotically degenerate, namely
it converges to the Dirac delta concentrated at $\rho_0$ if $k=0$ and at 0 if $k\neq 0$. From their
analysis it follows that $\xi_1=O(N)$ is possible
only if $k_1=0$. Moreover, let
\be {Aleps}
A_{L,\epsilon}=\left\{K\in(\Lambda^*)^N\left|\right.|N_0(K)/L^d-\rho_0|<\epsilon, N_k(K)/L^d<\epsilon
\mbox{ for } k\neq 0\right\},
\ee
then
\be{Paleps}
\lim_{L\to\infty} P_{\Lambda,N}(A_{L,\epsilon})=1 \mbox{ for any } \epsilon>0\ .
\ee
We prove equation (34) by showing that the probability of the complement of $A_{L,\epsilon}$
goes to zero.
\[
P_{\Lambda,N}(A_{L,\epsilon}^c)\leq P_{\Lambda,N}(A_{L,\epsilon}(k=0)^c)+
\sum_{k\neq 0}P_{\Lambda,N}(A_{L,\epsilon}(k)^c)
\]
where
\[
A_{L,\epsilon}(k=0)=\{K:|N_0(K)/L^d-\rho_0|<\epsilon\}
\]
and for $k\neq 0$
\[
A_{L,\epsilon}(k)=\{K:N_k(K)/L^d<\epsilon\}.
\]
Now $P_{\Lambda,N}(A_{L,\epsilon}(k=0)^c)$ goes to zero as $L$ tends to infinity. With
$
x_k=e^{-\beta\varepsilon(k)}
$
\begin{eqnarray}
P_{\Lambda,N}(A_{L,\epsilon}(k)^c)=
\frac{\sum_{m=\lceil\epsilon L^d\rceil}^N x_k^m
\sum_{\{n_q\}_{q\neq k}:\sum n_q=N-m}\prod_{q\neq k}x_q^{n_q}}
{\sum_{\{n_q\}:\sum n_q=N}\prod_{q}x_q^{n_q}}\nonumber\\
\leq x_k^{\epsilon L^d}
\frac{\sum_{m=\lceil\epsilon L^d\rceil}^N
\sum_{\{n_q\}_{q\neq k}:\sum n_q=N-m}\prod_{q\neq k}x_q^{n_q}}
{\sum_{\{n_q\}:\sum n_q=N}\prod_{q}x_q^{n_q}}\leq x_k^{\epsilon L^d}\nonumber
\end{eqnarray}
because in the numerator we have a part of the sum of the denominator, over those sets of occupation
numbers for which $\sum n_q=N$, $n_k=0$ and $n_0\geq \lceil\epsilon L^d\rceil$ (observe that $x_0=1$). But
\[
\beta\varepsilon(k)=\frac{\beta\hbar^2k^2}{2m}=\pi\lambda_B^2 n^2/L^2
\]
where $n=(n_1,\ldots,n_d)$, $n^2=\sum_{i=1}^d n_i^2$ with $n_i$ integers.
Therefore
\begin{eqnarray}
\sum_{k\neq 0}P_{\Lambda,N}(A_{L,\epsilon}(k)^c)&\leq&
\sum_{n\in Z^d\setminus\{0\}}\exp\{-\pi\epsilon\lambda_B^2 L^{d-2}n^2\}\nonumber\\
&=&\left(\sum_{i=-\infty}^\infty\exp\{-\pi\epsilon\lambda_B^2 L^{d-2}i^2\}\right)^d-1\nonumber\\
&=&\left[1+2\sum_{i=1}^\infty\exp\{-\pi\epsilon\lambda_B^2 L^{d-2}i^2\}\right]^d-1\nonumber\\
&\leq& \left[1+(\pi\epsilon\lambda_B^2 L^{d-2})^{-1/2}\right]^d-1\to 0\nonumber
\end{eqnarray}
with $L$ going to infinity if $d>2$.

The crucial property of the probabilities (\ref{probkg})
is that they are the same for all permutations leaving $K$ invariant.
Due to this uniform distribution and to (\ref{Paleps}), the discussion of the ground state cycle
percolation in I becomes relevant and we can derive the analogues of equations (I.5.16) and (I.5.18)
(notice the different meaning of $d$ -- the number of spin states -- in I). Related classical results
can also be found in Feller 1968.

We fix $\epsilon<\rho_0/2$ and choose a $K$ in $A_{L,\epsilon}$. Then $N_0(K)>N_k(K)$ for every $k\neq 0$.
We consider a $j$ such that $N_0(K)\geq j>\max_{k\neq 0} N_k(K)$.
Thus, cycles of length $j$ can occur only among particles with wavevector 0.
The conditional expectation value of $n_j$, provided that $K$ is given, is
\be{njcond}
\langle n_j\rangle_K\equiv \sum_m m P_{\Lambda,N}(n_j=m|K)={\sum_m m|\{g\in S_N|gK=K,n_j=m\}|\over
|\{g\in S_N|gK=K\}|}\ .
\ee
Here $|\{\cdots\}|$ means the number of elements of $\{\cdots\}$. Now the number of permutations
leaving $K$ invariant is $\prod_k N_k(K)!$, so we obtain
\be{nj2}
\langle n_j\rangle_K=\langle n_j\rangle_{N_0(K)}\equiv{1\over N_0(K)!}\sum_m m|\{g\in S_{N_0(K)}|n_j=m\}|
={1\over j}\ .
\ee
Indeed, for any $M\geq j$ the uniform
distribution $P_M(g)=1/M!$ over $S_M$ gives rise to
\be{njm}
\langle n_j\rangle_M=\sum_m mP_M(n_j=m)=\frac{1}{M!}\sum_m m|\{g\in S_M|n_j=m\}|=\frac{1}{j}\ ;
\ee
in order to see this it suffices to recall that the probability of a conjugation class $\{n_i\}$ of $S_M$
($\sum_i in_i=M$) is
\be{conj}
P_M(\{n_i\})=\prod_{i=1}^M {1\over n_i!}\left(\frac{1}{i}\right)^{n_i}\ .
\ee
From (\ref{nj2}) we obtain
\be{xicond}
P_{\Lambda,N}(\xi_1=j|K)={j\langle n_j\rangle_{N_0(K)}\over N_0(K)}\delta_{k_1,0}={\delta_{k_1,0}\over N_0(K)}\ .
\ee
Only particles with the same wave vector can be in the same cycle. Therefore,
$k_1=0$ implies $\xi_1\leq N_0(K)$ and
\be{pj}
P_{\Lambda,N}(\xi_1>j|K)=\left(1-{j\over N_0(K)}\right)\delta_{k_1,0}\ .
\ee
Fix now $0<x\leq\rho_0/\rho$. Equations (\ref{Paleps}) and (\ref{pj}) imply
\be{Px}
\lim_{L\to\infty} P_{\Lambda,N}\left(\frac{\xi_1}{N}>x\right)=\frac{\rho_0}{\rho}-x\ .
\ee
Indeed, by choosing $\epsilon<x\rho$,
\bea{limPx}
\lim_{L\to\infty}P_{\Lambda,N}\left(\frac{\xi_1}{N}>x\right)&=&
\lim_{L\to\infty}\sum_{K\in A_{L,\epsilon}}P_{\Lambda,N}\left(\frac{\xi_1}{N}>x|K\right)P_{\Lambda,N}(K)\nonumber\\
&=&\lim_{L\to\infty}\sum_{K\in A_{L,\epsilon}}\left(1-{xN\over N_0(K)}\right)\delta_{k_1,0}P_{\Lambda,N}(K)
\nonumber\\
&\lessgtr& \left(1-{x\rho\over \rho_0\pm\epsilon}\right)\lim_{L\to\infty}\sum_{K\in A_{L,\epsilon}}
\delta_{k_1,0}P_{\Lambda,N}(K)\nonumber\\
&=&\left(1-{x\rho\over \rho_0\pm\epsilon}\right)P_\rho(k_1=0)\nonumber\\
&=&\left(1-{x\rho\over \rho_0\pm\epsilon}\right)\frac{\rho_0}{\rho}\ .
\eea
The upper and lower bounds holding for all $\epsilon<x\rho$, we can send $\epsilon$ to zero and
find (\ref{Px}).

Complementary to this result is
\be{xi1j}
P_\rho(\xi_1=j|k_1=0)\equiv\lim_{L\to\infty}P_{\Lambda,N}(\xi_1=j|k_1=0)=0 \quad\mbox{for any fixed } j
\ee
whose analogue also existed in the ground-state cycle percolation, see (I.5.13). Equation
(\ref{xi1j}) can be obtained with the following argument.
If we restrict the probability
distribution $P_{\Lambda,N}(g|K)$ to the permutations of the $N_0(K)$
particles with 0 wavevector, the restricted distribution is still uniform. Therefore, by (\ref{njm}),
equation (\ref{nj2}) extends to cycles of an arbitrary length among the zero wavevectors.
As a consequence, if $K\in A_{L,\epsilon}$ is chosen so that $k_1=0$,
equation (\ref{xicond}) also extends to all $j\leq N_0(K)$
and together with (\ref{Paleps}) implies (\ref{xi1j}).
Thus, the full probability (\ref{Prho=}) comes from $k_1\neq 0$.

The number of infinite cycles of positive density is (countable) infinite.
The number of those with density larger than
$x\leq{\rho_0\over\rho}$ can be
obtained from equations (\ref{Paleps}) and (\ref{nj2}):
\be{sumnj}
\lim_{L\to\infty} \sum_{j=\lceil xN\rceil}^{N_0}\langle n_j\rangle_{\Lambda,N}=\ln\frac{\rho_0}{\rho x}\ .
\ee
This number can be arbitrarily large if $x$ is sufficiently small. For
$m>\ln\rho-\ln\rho_0$ the expected number
of cycles of density between $e^{-(m+1)}$ and $e^{-m}$ is
\[
\ln\frac{e^{m+1}\rho_0}{\rho}-\ln\frac{e^m\rho_0}{\rho}=\ln e=1.
\]
The intervals $[e^{-(m+1)},e^{-m})$ are disjoint, their number is infinite and on average there belongs
one infinite cycle to each interval.

Finally, we note that the above results apply without any further ado
to the imperfect Bose gas (cf Huang 1987). The Hamiltonian of this latter reads
\be{imper}
H_{\Lambda,N}=T_{\Lambda,N}+\frac{a}{2L^d}N^2
\ee
with some $a>0$, so probabilities and averages in the canonical ensemble coincide with those of the
perfect Bose gas.

\vspace{5pt}
\noindent {\bf Acknowledgment}
This work was supported by the Hungarian Scientific Research Fund under grant no T 30543.

\vspace{20pt}
\noindent
{\bf\large References}
\vspace{10pt}\\
Buffet E and Pul\`e J V 1983 {\it J. Math. Phys.} {\bf 24} 1608\\
Bund S and Schakel A M J 1999 {\it Mod. Phys. Lett.} B {\bf 13} 349\\
Cannon J T 1973 {\it Commun. Math. Phys.} {\bf 29} 89\\
Feller W 1968 {\it An introduction to probability theory and its applications I.}\\
--------- (New York: John Wiley) Ch. X. 6\\
Feynman R P 1953 {\it Phys. Rev.} {\bf 91} 1291\\
Huang K 1987 {\it Statistical Mechanics} (New York: Wiley)\\
Lewis J T and Pul\`e J V {\it Commun. Math. Phys.} {\bf 36} 1\\
Martin Ph 2001 private communication\\
Schakel A M J 2001 {\it Phys. Rev.} E {\bf 63} {026115}\\
S\"ut\H o A 1993 {\it J. Phys. A: Math. Gen.} {\bf 26} 4689\\
Ueltschi D 2002 {\it In and out of equilibrium: Physics with a probability flavor}\\
----------- ed V Sidoravicius
(Basel: Birkh\"auser) p 363

\end{document}